\documentclass[aps,prl,showpacs,amsmath,amssymb,amsfonts,lengthcheck,twocolumn,longbibliography,superscriptaddress]{revtex4-2}

\usepackage[T1]{fontenc}
\usepackage{graphicx}
\usepackage{physics}
\usepackage{subfigure}
\usepackage{amsthm}
\usepackage{amsmath}
\usepackage{verbatim}
\usepackage{dcolumn}
\usepackage{bm}
\usepackage{epsf}
\usepackage{color}
\usepackage[colorlinks=true,citecolor=blue,linkcolor=blue,urlcolor=blue]{hyperref}%
\usepackage{xcolor}
\usepackage{dsfont}
\usepackage{tikz}
\usepackage{pgfplots}
\usepackage{xcolor}
\usepackage{url}
\usepackage{placeins}
\usepackage[normalem]{ulem}

\definecolor{myvermillion}{HTML}{D55E00}
\definecolor{myyellow}{HTML}{F0E442}
\definecolor{mygreen}{HTML}{009E73}
\definecolor{mypurple}{HTML}{CC79A7}
\definecolor{myblue}{HTML}{0072B2}
\definecolor{myorange}{HTML}{E69F00}
\definecolor{mygray}{RGB}{153,153,153}

\pgfplotsset{compat=newest}

\newcommand{\bla}{bla\\bla\\bla\\bla\\bla}

\makeatletter
\newcommand{\currentfontsize}{\f@size pt}
\makeatother

\usepackage[colorlinks=true,citecolor=blue,linkcolor=blue,urlcolor=blue]{hyperref}

\begin{document}

\title{Quantum speed limit for the OTOC from an open systems perspective}

\author{Devjyoti Tripathy}
\email{dtripathy@umbc.edu}
    \affiliation{Department of Physics, University of Maryland, Baltimore County, Baltimore, MD 21250, USA}
    \affiliation{Quantum Science Institute, University of Maryland, Baltimore County, Baltimore, MD 21250, USA}

\author{Juzar Thingna}
\affiliation{American Physical Society, 100 Motor Parkway, Hauppauge, New York 11788, USA}
\affiliation{Center for Theoretical Physics of Complex Systems, Institute for Basic Science (IBS), Daejeon 34126, Republic of Korea}

\author{Sebastian Deffner}
    \thanks{JT and SD contributed equally to this work}
    \affiliation{Department of Physics, University of Maryland, Baltimore County, Baltimore, MD 21250, USA}
    \affiliation{Quantum Science Institute, University of Maryland, Baltimore County, Baltimore, MD 21250, USA}
    \affiliation{National Quantum Laboratory, College Park, MD 20740, USA}
\date{\today}

\begin{abstract}
    Scrambling, the delocalization of initially localized quantum information, is commonly characterized by the out-of-time ordered correlator (OTOC). Employing the OTOC–Renyi-2 entropy theorem we derive a quantum speed limit for the OTOC, which sets an lower bound for the rate with which information can be scrambled. This bound becomes particularly tractable by describing the scrambling of information in a closed quantum system as an effective decoherence process of an \emph{open} system interacting with an environment. We prove that decay of the OTOC can be bounded by the strength of the system-environment coupling and two-point environmental correlation functions. We validate our analytic bound numerically using the non-integrable transverse field Ising model. Our results provide a universal and model-agnostic quantitative framework for understanding the dynamical limits of information spreading across quantum many-body physics, condensed matter, and engineered quantum platforms.
\end{abstract}

\maketitle


\paragraph{Introduction}
One of the most profound features that distinguishes quantum mechanics from classical physics is the non-commutativity of observables, which underpins the Heisenberg uncertainty relations \cite{1927ZPhy...43..172H,PhysRev.34.163}. A textbook interpretation of these relations is that the precise value of two non-commuting observables cannot be determined in simultaneous measurements \cite{Messiah2014quantum}. The correlation functions of such non-commuting observables then quantify how much information ``can be found out'' about one observable from measuring the other. Arguably, one of the most important applications of this fundamental insight is Hahn's spin echo \cite{HahnPR1950}, in which context also the out-of-time ordered correlator (OTOC) was defined \cite{Swingle2018NP, Roberts2016PRL}. 

The OTOC is a four-point correlation function that captures how the commutativity of two local observables evolves in a closed quantum system. It can be written as
\begin{equation}
\label{eq:otoc4pt}
    \mathcal{O}(t)=\langle B^{\dagger}(t)A^{\dagger}B(t)A\rangle_{\rho},
\end{equation}
where $A$ and $B$ are local operators acting on subsystems $\mathfrak{A}$ and $\mathfrak{B}$, and $\rho$ is the initial state of the composite system $\mathfrak{A}\otimes\mathfrak{B}$. The operator $B(t)=\exp{iHt}\,B\,\exp{-iHt}$ represents the time-evolution of $B$ in the Heisenberg picture under the Hamiltonian of the total system $H$, and we set $\hbar=1$. For initially commuting operators $A$ and $B$, the decay of $\mathcal{O}(t)$ signals the growth of operator $B(t)$ into the support of $A$.

In recent years, the OTOC \eqref{eq:otoc4pt} has found widespread application \cite{Swingle2018NP,Huang2017AP,Zhu2022PRL,Landsman2019,Gärttner2017,PhysRevA.95.012120} in the context of quantum information scrambling \cite{Touil2024EPL,Touil2021PRX,Roberts2016PRL,Anand2022brotocsquantum,Dong2022PRR,PatrickHayden_2007,Blok2021PRX,Zanardi2024operationalquantum,Dallas2024Butterfly,PhysRevA.109.052424,PhysRevA.107.042217,Zanardi2022quantumscramblingof,PhysRevA.103.062214,PhysRevA.110.052416,MatsoukasRoubeas2024quantumchaos,PhysRevLett.131.160202}.  Quantum information scrambling refers to the spread of initially localized quantum information throughout non-local degrees of freedom under unitary dynamics, becoming effectively inaccessible to local measurements \cite{Lo_Monaco_2025}. The actual information is hidden in correlations between different subsystems and can only be recovered by global measurements. This delocalization is central to understanding thermalization and chaos in complex quantum systems such as spin chains or black holes \cite{Hosur2016,Shenker2014,Patrick_Hayden_2007,PhysRevE.102.022201,PRXQuantum.5.010201,PhysRevLett.124.200504,PhysRevB.108.134305,PhysRevE.81.017203,PhysRevResearch.7.013146,PhysRevResearch.7.013181,PhysRevB.109.224304,PhysRevResearch.4.033093,PhysRevB.103.064309,Chenu2019workstatistics,PhysRevLett.122.014103,Chenu2018,PhysRevResearch.7.L022027,Carolan_2024}. Interestingly, in quantumly chaotic systems \cite{10.1063/5.0199335} the OTOC takes a particularly simple from $\mathcal{O}(t)\sim \exp{-\lambda \,t}$, where $\lambda$ is called  quantum Lyapunov exponent \cite{Hallam2019,PhysRevE.101.010202,Maldacena2016,PhysRevD.110.086010}.

More generally, computing the OTOC is an involved problem as it requires the solution of the complex quantum many-body dynamics. In the present letter, we show that employing the framework of quantum speed limits (QSL) the problem can be made tractable, even for quantumly chaotic dynamics. QSLs are a careful reformulation of Heisenberg's uncertainty relation for energy and time, and they provide tight bounds on the rate with which quantum states can evolve \cite{Mandelstam1991}. In particular, with their generalization to open system dynamics \cite{Funo_2019,PhysRevResearch.1.033127,PhysRevLett.111.010402,PhysRevLett.110.050403}, one of the recently most active fields of research emerged. In rapid succession QSLs were generalized to mixed states \cite{PhysRevA.67.052109}, time-dependent Hamiltonians \cite{Deffner_2013, UHLMANN1992329, PhysRevLett.70.3365}, and and applied to many other quantities such as measures of entanglement \cite{PhysRevA.67.052109, PhysRevA.72.032337, PhysRevA.104.032417} and coherence \cite{Mohan2022NJP}, quantum observables \cite{PhysRevResearch.6.013018,PhysRevA.106.042436,AiferPRL24} and complexity \cite{Hörnedal2022,gill2024speedlimitsscramblingkrylov, PhysRevD.109.L121902}, and many other problems \cite{Lloyd2000,PhysRevLett.130.010402,PRXQuantum.2.040349, Frey2016, Aifer_2022,SilvaPratapsi_2025,Deffner_2022,Aifer_2022,Deffner2021quantumspeedlimite,PhysRevResearch.2.013161,PhysRevLett.111.010402}.

Our main result is a QSL for the OTOC, which can be leveraged, e.g., to assess how quickly quantum information is scrambled. Interestingly, some of us have found in previous work that quantum information scrambling can be effectively described by methods of decoherence theory \cite{10.1063/5.0199335}. This allows to express our novel QSL for the OTOC to be expressed in terms of a two-point correlator of the reduced state. This approach is akin to Kubo's formula \cite{Kubo:1957mj} expressing transport coefficients via linear response functions. Importantly, our bounds are experimentally much more accessible than the OTOC itself, since two-point functions are significantly easier to measure \cite{bocini2024nonlocalquenchspectroscopyfermionic,Schweigler2017,Zhang2024,Islam2015,PhysRevLett.127.200501,PhysRevD.105.074503} than four-point correlators. Moreover, our QSL recasts the many-body system's degrees of freedom as an effective environment and serves as a tool to classify how environmental correlations influence the rate of scrambling. This is illustrated for the non-integrable transverse field Ising chain, which demonstrates that our approach offers both clarity and practical utility in quantifying the limits of quantum information delocalization.


\paragraph{Bounding the OTOCs}


While OTOCs can provide deep insights into the spreading of information and the growth of operators and entanglement \cite{XuPRA19,PhysRevX.8.021013,Carolan_2024,PhysRevLett.126.030601,Anand2022brotocsquantum}, they pose experimental and computational challenges \cite{PhysRevLett.128.140601,doi:10.1126/science.abg5029,Braumüller2022} due to their nontrivial temporal ordering. To address this issue, we recast scrambling within a broader framework of \emph{open} quantum systems. By treating a subsystem of interest as evolving under the interaction of its many-body environment, we reinterpret scrambling as a form of decoherence. As a main result, we derive a fundamental lower bound on the decay of the $\mathcal{O}(t)$, thus constraining how fast information can scramble in a many-body quantum system.

We start with the OTOC–Renyi-2 entropy theorem \cite{FAN2017707,zhou2025measuringrenyientropyusing}. Consider a system with an initial state $\rho(0)$. After being quenched by an arbitrary operator $O$ at $t=0$, we divide the system into two subsystems $\mathfrak{A}$ and $\mathfrak{B}$. The quench allows us to apply a local perturbation to subsystem $\mathfrak{A}$, which then quickly decoheres into subsystem $\mathfrak{B}$. In the language of open quantum systems, subsystem $\mathfrak{B}$ is referred to as an environment. Then the average OTOC on the composite system is related to purity of subsystem $\mathfrak{A}$ \cite{FAN2017707,zhou2025measuringrenyientropyusing},
\begin{equation}
    \label{eq:otoc-re}
    \Bar{\mathcal{O}}(t) = \exp{-S_\mathfrak{A}^{(2)}(t)}\,,
\end{equation}
where
\begin{equation}
 S_\mathfrak{A}^{(2)}(t) = -\ln\left(\tr\left\{\rho_\mathfrak{A}^2(t)\right\}\right)
\end{equation} 
is the Renyi-2 entropy of subsystem $\mathfrak{A}$. The average OTOC, $\Bar{\mathcal{O}}(t)$ is defined as a Haar average over all unitaries on subsystem $\mathfrak{B}$,i.e.,
\begin{equation}
    \Bar{\mathcal{O}}(t) = \int dB \, \tr\left\{B^{\dagger}(t)A^{\dagger}B(t)A\right\},
\end{equation}
and $A = O\rho(0)O^{\dagger}$. Equation~(\ref{eq:otoc-re}) requires the full density matrix and thus obtaining the OTOC via the R.H.S. of Eq.~(\ref{eq:otoc-re}) is as hard as the OTOC itself. Our goal now has to be to find a simpler bound that depends merely on computing two-point environmental correlators that are tractable.

The Hamiltonian of the global quantum system reads,
\begin{equation}
    H = H_\mathfrak{A} + H_\mathfrak{B} + \lambda H_I\quad \text{and} \quad H_I = \sum_\alpha \mathcal{A}_\alpha \otimes \mathcal{B}_\alpha,
\end{equation}
where $H_{\mathfrak{A}}$, $H_{\mathfrak{B}}$ are the subsystem Hamiltonians, respectively, and $H_{I}$ is the interaction Hamiltonian between $\mathfrak{A}$ and $\mathfrak{B}$ with unitless interaction strength $\lambda$. The reduced state $\rho_\mathfrak{A}(t) = \tr_\mathfrak{B}\{\rho(t)\}$ evolves non-unitarily, acquiring entropy as it entangles with subsystem $\mathfrak{B}$.

In the interaction picture, the dynamics can be written as,
\begin{equation}
    \dot{\rho}(t) = -i\lambda\, [H_I(t), \rho(t)]\,,
\end{equation}
where, as always, the dot denotes a derivative with respect to time. Expanding up to second order in $\lambda$ and tracing over subsystem $\mathfrak{B}$~\cite{ThingnaJCP12,BeckerPRL22}, gives,
\begin{equation}
   \dot{\rho}_\mathfrak{A}(t) = -\lambda^2 \int_0^t ds \, \tr_\mathfrak{B} \left\{ H_I(t), [H_I(s), \rho(s)] \right\}.
\end{equation}
Up to this point, we have made no assumptions about the size or properties of the subsystem $\mathfrak{B}$. Now, we assume that the subsystem $\mathfrak{B}$ is large enough such that the dynamics of $\mathfrak{A}$ does not affect it. In other words, there is no backaction of $\mathfrak{A}$ on $\mathfrak{B}$. Thus, we can decompose the total density matrix $\rho(s)=\rho_{\mathfrak{A}}(s)\otimes\rho_{\mathfrak{B}}^{\beta}$. Consistent and motivated by arguments of canonical typicality \cite{Goldstein2006PRL} we further assume $\rho_{\mathfrak{B}}^{\beta}=\exp{-\beta H_{\mathfrak{B}}}/Z$, which is invariant under the total evolution. This is also commonly referred to as the Born approximation in the theory of open quantum systems~\cite{10.1093/acprof:oso/9780199213900.001.0001, ThingnaJCP14}

Under the weak coupling approximation, $\lambda\ll\tau_{\mathfrak{B}}^{-1}$ where $\tau_{\mathfrak{B}}$ is the timescale for the decay of environmental correlations, one then can derive the Redfield master equation \cite{Redfield1965},
\begin{equation}
    \dot{\rho}_{\mathfrak{A}}(t) = -\lambda^{2}\int_{0}^{t}ds\tr_{\mathfrak{B}}\left\{H_{I}(t),[H_{I}(s),\rho_{\mathfrak{A}}(t)\otimes\rho_{\mathfrak{B}}^{\beta}]\right\}.
\end{equation}
Assuming a general form of the interaction Hamiltonian $H_{I}=\sum_{i}\mathcal{A}_{i}\otimes \mathcal{B}_{i}$, we obtain,
\begin{align}
\label{eq:redfield qsl}
\dot{\rho}_\mathfrak{A}(t) 
=& -\lambda^2 \int_0^t ds \sum_{i,j} \Gamma_{i,j}(t,s)\,
   \Bigl(\mathcal{A}_{i}(t
   )\mathcal{A}_{j}(s)\rho_{\mathfrak{A}}(t) \notag \\
 &  - \mathcal{A}_{j}(s)\rho_{\mathfrak{A}}(t)\mathcal{A}_{i}(t
   )\Bigl)+ \text{h.c.} 
\end{align}
where $\Gamma_{i,j}= \tr_B \left\{\mathcal{B}_i(t) \mathcal{B}_j(s)\rho_{\mathfrak{B}}^{\beta}\right\}$ and $\mathcal{A}_{i}(t)$ and $\mathcal{B}_{j}(t)$ evolved according to their respective subsystem Hamiltonians.


Having recast scrambling as loss of purity in an open system, we now turn to bounding the rate of this purity loss. The dynamics of $\rho_\mathfrak{A}(t)$ given by Eq.~(\ref{eq:redfield qsl}) can be written abstractly as, $\dot{\rho}_\mathfrak{A}(t) = \mathcal{L}_t[\rho_\mathfrak{A}(t)]$, where $\mathcal{L}_t$ is the time-dependent Liouvillian superoperator. Then, the Renyi-2 entropy of subsystem $\mathfrak{A}$ satisfies \cite{Uzdin_2016},
\begin{equation}
\label{eq:qsl-state}
    S_\mathfrak{A}^{(2)}(t) \leq 2 \int_0^t dt'\, \| \mathcal{L}_{t'}(\rho_\mathfrak{A}) \|_{\text{sp}} .
\end{equation}
 where $\norm{.}_\text{sp}$ is the spectral operator norm. A tighter bound is obtained in Liouville space, where the density matrix is vectorized $\ket{\rho_\mathfrak{A}}$. The equation of motion in Liouville space is given by, $i\ket{\dot{\rho}_\mathfrak{A}} = L_t \ket{\rho_\mathfrak{A}}$, and the corresponding bound becomes \cite{Uzdin_2016},
\begin{equation}
\label{eq:qsl-liouville}
    S_\mathfrak{A}^{(2)}(t) \leq \int_0^t dt'\,\| L_{t'} - L_{t'}^\dagger \|_{\text{sp}} .
\end{equation}

Using Eq.~\eqref{eq:otoc-re}, which relates the OTOC to the Renyi-2 entropy, and applying the entropy bounds Eqs.~\eqref{eq:qsl-state} and \eqref{eq:qsl-liouville}, we obtain our main analytic result, namely, the desired QSL for the OTOC,
\begin{subequations}
\label{eq:qsl-final}
\begin{align}
    \Bar{\mathcal{O}}(t) &\geq \exp{-\int_0^t dt'\, \| L_{t'} - L_{t'}^\dagger \|_{\text{sp}}},
    \label{eq:qsl-final-a} \\
    &\geq \exp{ -2 \int_0^t dt'\,\| \mathcal{L}_{t'}(\rho_\mathfrak{A}) \|_{\text{sp}}}.
    \label{eq:qsl-final-b}
\end{align}
\end{subequations}
These bounds provide a direct constraint on the decay of the OTOC and hence on the rate of information scrambling in quantum systems. 

Before we continue with demonstrating the utility of our bounds \eqref{eq:qsl-final} in a specific example, we briefly remark on the chaotic limit. In situation in which the OTOC becomes exponential, $\mathcal{O}(t)\sim \exp{-\lambda \,t}$, Eqs.~\eqref{eq:qsl-final} yield 
\begin{equation}
\label{eq:lya}
    \lambda \leq \int_0^t dt'\, \| L_{t'} - L_{t'}^\dagger \|_{\text{sp}}\leq 2 \int_0^t dt'\,\| \mathcal{L}_{t'}(\rho_\mathfrak{A}) \|_{\text{sp}}.
\end{equation}
In other words, our QSL for the OTOC \eqref{eq:qsl-final} also provides a (tight) upper bound on the quantum Lyapunov exponents in terms of environmental 2-point correlation functions.

In the next section, we now demonstrate the utility of results for the non-integrable transverse field Ising model.



\paragraph{Ferromagnetic Non-integrable Transverse Field Ising Model}

\begin{figure}
    \includegraphics[width=0.48\textwidth]{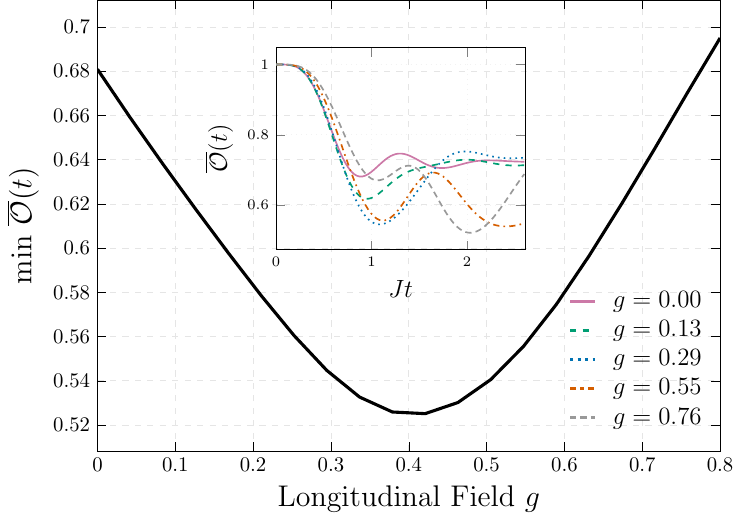}
    \caption{Minimum value of $\Bar{\mathcal{O}}(t)$ for $Jt\in [0,1.2]$ with increasing magnitude of the longitudinal field $g$ for the ferromagnetic non-integrable transverse field Ising model with $N=10$, $J=0.65$, $h=0.9$, $\beta=1.0$. The inset shows the behavior of $\Bar{\mathcal{O}}(t)$ with time for various values of the longitudinal field $g$.} 
    \label{fig:min_renyi}
\end{figure}

The one-dimensional transverse field Ising Hamiltonian with a longitudinal field and open boundary conditions reads \cite{PhysRevB.92.104306,PhysRevE.90.052105,PhysRevB.106.214311},
\begin{equation}
\label{eq:fm_tfim}
    H = -J \sum_{i=1}^{N-1} \sigma_{i}^{z} \sigma_{i+1}^{z} - h \sum_{i=1}^{N} \sigma_{i}^{x} - g \sum_{i=1}^{N} \sigma_{i}^{z}.
\end{equation}
The chain has $N$ spin-$1/2$ nearest-neighbor interacting particles with Pauli matrices $\sigma_{i}^{\alpha}$, where $\alpha = x,y,x$ is the $\alpha$th component at site $i$. When $g=0$, the Hamiltonian reduces to the usual quantum Ising Model in the transverse field \cite{PFEUTY197079, PhysRevLett.110.135704} and becomes integrable using the Jordan-Wigner transformation. At $h=0$, it has a quantum critical point. For $h\neq0$ and $g\neq0$, the model becomes non-integrable. For $J >0$, the Ising coupling energetically favors ferromagnetic ordering.

We consider subsystem $\mathfrak{A}$ to consist of just the first spin at site 1. Subsystem $\mathfrak{B}$ consists of the rest $(N-1)$ spins, which act as a finite-sized environment for $\mathfrak{A}$. Thus, we decompose the chain into subsystems $\mathfrak{A}$ and $\mathfrak{B}$ as,
\begin{equation}
H_{\mathfrak{A}} = -h \sigma_{1}^{x} - g\sigma_{1}^{z},
\end{equation}
and for the ``environment'' $\mathfrak{B}$ we have,
\begin{equation}
H_{\mathfrak{B}} = -J \sum_{i=2}^{N-1} \sigma_{i}^{z} \sigma_{i+1}^{z} - h \sum_{i=2}^{N}
    \sigma_{i}^{x} - g\sum_{i=2}^{N}\sigma_{i}^{z}.
\end{equation}
Finally, the interaction term is given by
\begin{equation}
H_{I} = -J \sigma_{1}^{z} \sigma_{2}^{z}.
\end{equation}

Subsystem $\mathfrak{A}$ is initialized in a pure state $\rho_\mathfrak{A} = \ket{0}\bra{0}$. We then quench the initial state with $O = \sigma_\mathfrak{A}^{x}$ such that $\rho(0)$ transforms to $O(\ket{0}\bra{0}\otimes\exp{-\beta H_{\mathfrak{B}}}/Z)O^\dagger$. Correspondingly, the  Redfield master equation for the non-integrable Ising model in interaction picture becomes,
\begin{widetext}
\begin{align}
\label{eq:tfim_redfield}
    \dot{\rho}_{\mathfrak{A}}(t) &= -J^{2}\int_{0}^{t}ds \Big[ \Gamma(t,s) \big( \sigma_{1}^{z}(t) \sigma_{1}^{z}(s) \rho_{\mathfrak{A}}(t) 
    - \sigma_{1}^{z}(s) \rho_{\mathfrak{A}}(t) \sigma_{1}^{z}(t) \big) + \Gamma(t,s)^{*} \big( \rho_{\mathfrak{A}}(t) \sigma_{1}^{z}(s) \sigma_{1}^{z}(t) 
    - \sigma_{1}^{z}(t) \rho_{\mathfrak{A}}(t) \sigma_{1}^{z}(s) \big) \Big]
\end{align}
\end{widetext}
where $\Gamma(t,s) = \langle \sigma_{2}^{z}(t) \sigma_{2}^{z}(s) \rangle_{\rho_{\mathfrak{B}}^{\beta}}$ is the two-point correlation function of subsystem $\mathfrak{B}$.

\begin{figure}
    \centering
    \includegraphics[scale=0.7]{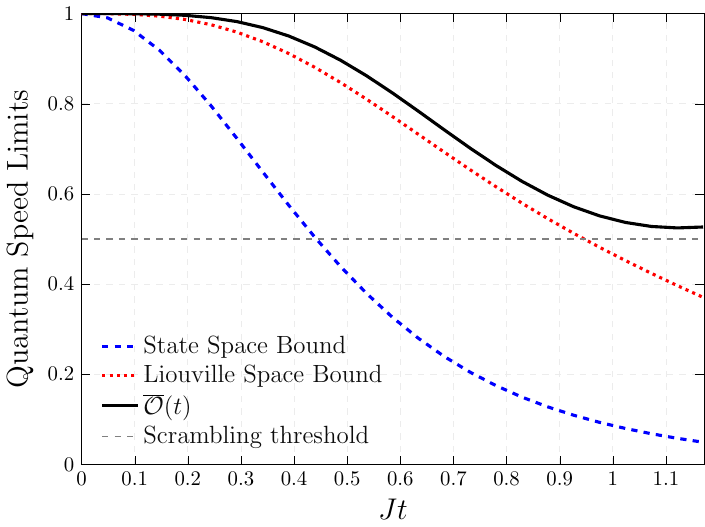}
    \caption{QSL for the ferromagnetic non-integrable transverse field Ising model with $N=10$, $J=0.65$, $h=0.9$, $g=0.4$, $\beta=1.0$.  The gray dashed line represents the minimum value for the OTOC $\Bar{\mathcal{O}}(t)$, when the reduced state $\rho_\mathfrak{A}$ is maximally mixed.}
    \label{fig:qsl_tfim}
\end{figure}

An important aspect of this analysis is the hierarchy of time scales inherent in an open system description. The first is the intrinsic time scale of the subsystem $\mathfrak{A}$, given by $\tau_{\mathfrak{A}}=(2\sqrt{h^{2}+g^{2}})^{-1}$. The second is the interaction time scale between the subsystem $\mathfrak{A}$ and subsystem $\mathfrak{B}$, $\tau_{I} = J^{-1}$ and third is the correlation decay timescale within subsystem $\mathfrak{B}$, $\tau_{\mathfrak{B}}$. 

For the Redfield master equation \eqref{eq:tfim_redfield} to be valid, the standard condition must be satisfied: $\tau_{\mathfrak{B}}\ll\tau_{I}\ll\tau_{\mathfrak{A}}$. Such a description of the master equation is traditionally discussed in the limit of an infinitely large environment ensuring a rapid correlation decay.

We observe that this separation of time scales also holds in our finite-sized numerics. More precisely, the chaotic nature of the subsystem $\mathfrak{B}$ effectively induces a small $\tau_{\mathfrak{B}}$. Moreover, as shown in Appendix A, our bounds Eq.~(\ref{eq:qsl-final}) and the early time behavior of the OTOCs coincide for different composite system sizes $N$, indicating the validity of Born approximation even in finite environments. 

To evaluate the bounds \eqref{eq:qsl-final} numerically, we compute the two point correlators using a Crank-Nicolson time evolution scheme. We then construct the Liouvillian, numerically integrate it, and substitute it into the bound expressions Eq.~(\ref{eq:qsl-final}). We emphasize that, in our setting, the Redfield equation \eqref{eq:tfim_redfield} is built with a finite subsystem $\mathfrak{B}$. Consequently, the bath two-point correlators $\Gamma(t,s)$ entering the master equation, as well as the $\Bar{\mathcal{O}}(t)$ of the full chain, inevitably exhibit finite-size revivals. To avoid late-time recurrences, we restrict all analyses to a pre-revival window, chosen up to the first minimum of $\bar{\mathcal O}(t)$. In this regime the bounds are tight.

To meaningfully characterize the onset of scrambling in this model, we extract the minimum value of the numerically exact OTOC $\bar{\mathcal{O}}(t)$ prior to the emergence of finite time oscillatory effects. This minimum is a coarse grained measure of the maximum delocalization that our model can achieve before partial revivals or late time correlations take over the dynamics. We plot this quantity in Fig.~\ref{fig:min_renyi} as a function of the longitudinal field strength $g$ which induces non-integrability in the model.
 
We find that the minimum value of $\bar{\mathcal{O}}(t)$, initially, decreases with increase of $g$, achieves a minimum at $g\approx0.4$ and then increases with increasing $g$. The inset in Fig.~\ref{fig:min_renyi} shows the time evolution of $\bar{\mathcal{O}}(t)$ for various values of $g$, clearly illustrating the sharp initial dip that precedes revival effects. This early time minimum, therefore, depicts the strength and irreversibility of information spreading in the system. Figure~\ref{fig:qsl_tfim} shows our analytic bounds,i.e., the QSL \eqref{eq:qsl-final}, for the non-integrable transverse field Ising model. The gray dashed line is the scrambling threshold, which corresponds to the lowest possible value of $\bar{\mathcal{O}}(t)|_{\textrm{min}} = 0.5$. This value is obtained when the Renyi entropy is maximized,i.e., when $\rho_\mathfrak{A}$ is maximally mixed. We also find that, as expected, the Liouville space bound Eq.~(\ref{eq:qsl-final-a}) is a significantly tighter bound, very close to the actual $\bar{\mathcal{O}}(t)$, than the state space bound Eq.~(\ref{eq:qsl-final-b}). Since we bound the $\bar{\mathcal{O}}(t)$ using only two-point correlators, it  can become looser than $\bar{\mathcal{O}}(t)|_{\textrm{min}}$. Our bounds, getting lower than 0.5, hence, do not indicate any unphysical behavior of the OTOC, only the bound is not tight in this regime. Moreover, we restrict our plots to $Jt\leq 1.2$, since by this point the finite-sized effects have taken over and the bounds have become loose.
 
\begin{figure}
    \centering
    \includegraphics[scale=0.7]{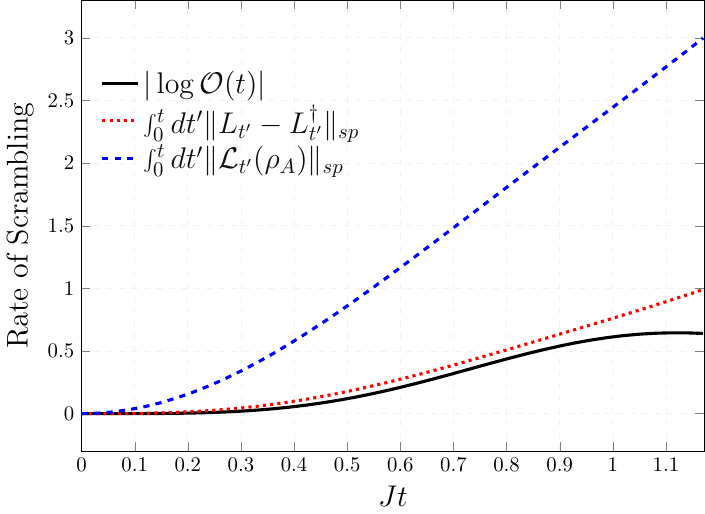}
    \caption{Rate of information scrambling in the ferromagnetic non-integrable transverse field Ising model with $N=10$, $J=0.65$, $h=0.9$, $g=0.4$, $\beta=1.0$. The actual rate of scrambling is given by the black curve. The red and blue curves are obtained from the absolute value of logarithm of the QSL bounds given by Eq.~(\ref{eq:qsl-final})}.
    \label{fig:rate of scrambling}
\end{figure}

Furthermore, to quantify the rate of scrambling, we examine the time evolution of $\abs{\log\Bar{\mathcal{O}}(t)}$ and compare it against the absolute value of the logarithm of the Liouville space and State space bounds. This is shown in Fig.~\ref{fig:rate of scrambling}. In systems and regimes that exhibit exponential scrambling, this quantity grows linearly in time, with the slope corresponding to the quantum Lyapunov exponent, cf. Eq.~\eqref{eq:lya}. Although our model here does not generically show such an exponential behavior ~\cite{PhysRevB.108.L121108}, plotting $\abs{\log\Bar{\mathcal{O}}(t)}$ enables us to obtain a characteristic rate of growth of scrambling during early-time dynamics.

In Appendix A, we show QSLs for the case when we relax the assumption that the subsystem $\mathfrak{B}$ is in a stationary thermal state at all times (Born approximation). Within the Liouville space, we find that the bound with a non-stationary subsystem $\mathfrak{B}$ is slightly tighter than the bound with the Born approximation. This should be expected since accounting for the dynamics of subsystem $\mathfrak{B}$ is more accurate, albeit at a significant computational cost.

\paragraph{Antiferromagnetic Non-Integrable Transverse Field Ising Model}

To conclude the analysis, we also solve the antiferromagnetic version of the non-integrable Ising model, whose Hamiltonian is Eq.~(\ref{eq:fm_tfim}) with $J <0$.

\begin{figure}
    \centering
    \includegraphics[scale=0.7]{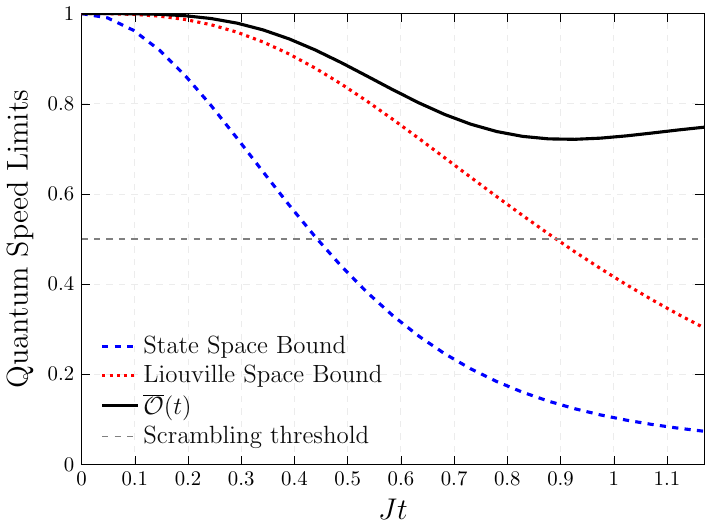} 
    \caption{QSL for the antiferromagnetic non-integrable transverse field Ising model with $N=10$, $J=0.65$, $h=0.9$, $g=0.4$, $\beta=1.0$.}
    \label{fig:qsl_afm_tfim}
\end{figure}

Figure~\ref{fig:qsl_afm_tfim} shows the OTOC and our QSL  \eqref{eq:qsl-final} for the antiferromagnetic case, with parameters identical to the ferromagnetic case. In comparing the scrambling behavior of the ferromagnetic and antiferromagnetic versions, a notable difference emerges in the extent to which the OTOC decays. Specifically, in the ferromagnetic case, the system exhibits significantly faster and more complete scrambling, with OTOCs decaying to lower values compared to the antiferromagnetic case. This can be attributed to the nature of the spin coupling: in the ferromagnetic case, the interaction term favors all spins aligning in the same direction, so a local perturbation rapidly destabilizes this uniform configuration and propagates freely through the chain. In contrast, the antiferromagnetic interaction energetically favors alternating spin alignment. As a result, a perturbation attempting to flip a single spin encounters immediate energetic resistance from its neighbors, leading to frustration and slowing the spread of information. This resistance inhibits the global delocalization of quantum information, thus suppressing the decay of the OTOC. Consequently, the effective timescale for scrambling is larger in the antiferromagnetic case, and the system retains a higher degree of local coherence over time. In Appendix B, we demonstrate a perturbative regime in which the QSLs can be obtained analytically.


\paragraph{Concluding remarks}

In the present letter, we have developed a rigorous framework for bounding information scrambling in quantum many-body systems by bounding the decay of purity and hence the OTOC, employing the framework of QSL. By relating the reduced dynamics in open quantum systems to scrambling in closed many-body systems and leveraging norm constructions, we derived bounds that estimate the rate of entanglement generation between a reduced subsystem and its complement. Crucially, we have demonstrated that the OTOC, a four-point function, is bounded below by products of two-point environment correlators and the spectral norm of system jump operators. This inequality reveals that scrambling in quantum systems is constrained by both the strength of coupling (via jump operators) and coherence of environmental correlations.  

Numerical simulations of the non-integrable transverse field Ising model confirm the tightness and qualitative behavior of these bounds. We have observed that the bounds closely track the early time dip of the OTOCs demonstrating the robustness of our bound in integrability breaking regimes. This QSL is thus a powerful framework to quantitatively constrain scrambling even in the presence of decoherence, explaining the interplay between dissipation, entanglement growth, and information spreading.

An important consequence of our work lies in exploiting the spectral structure of the environment (referred to as subsystem $\mathfrak{B}$ above), such as Ohmic, sub-ohmic or super-Ohmic forms \cite{PhysRevB.92.195143}. In these cases, the bounds can be computed analytically or semi-analytically, drastically reducing the computational complexity. This opens up the possibility of classifying chaotic unitaries by their environment's effective noise spectrum, thereby building up the theoretical framework of dissipative quantum chaos~\cite{wold2025experimentaldetectiondissipativequantum,sá2023signaturesdissipativequantumchaos,PhysRevResearch.7.013276,fritzsch2025freecumulantseigenstatethermalization}.

Moreover, our results may lead to experimental investigation of scrambling in platforms where the coupling to the environment can be engineered \cite{MIRRAHIMI2013424} such as trapped ions \cite{ PhysRevLett.77.4728, Andrade_2022}, superconducting qubits \cite{PhysRevResearch.1.013004, Kitzman2023,hu2024engineeringenvironmentsuperconductingqubit, MIRRAHIMI2013424}, and cold atoms \cite{schönleber2017reservoirengineeringultracoldrydberg, Schönleber_2018, Xie_2022} coupled to engineered reservoirs.

\paragraph{Data Availability} The code for our numerical simulation can be found at \cite{github}.

\acknowledgments{S.D. acknowledges support from the John Templeton Foundation under Grant No. 62422. This work was supported by the U.S. Department of Energy, Office of Basic Energy Sciences, Quantum Information Science program in Chemical Sciences, Geosciences, and Biosciences, under Award No. DE-SC0025997.}


\bibliography{ref}

\clearpage
\onecolumngrid
\begin{center}
    \textbf{\Large End Matter}
\end{center}
\vspace{1em}
\twocolumngrid

\setcounter{section}{0}
\renewcommand{\thesection}{\arabic{section}}
\renewcommand{\theequation}{\arabic{equation}}
\renewcommand{\thefigure}{\arabic{figure}}

\paragraph{Appendix A: Ferromagnetic Transverse Field Ising Model}

In the main text, we derive the Redfield master equation \eqref{eq:tfim_redfield} for the non-integrable transverse field Ising model with ferromagnetic coupling,
\begin{widetext}
\begin{align}
    \dot{ \rho_{\mathfrak{A}}} &= -J^{2}\int_{0}^{t}ds \Big[ \Gamma(t,s) \big( \sigma_{1}^{z}(t) \sigma_{1}^{z}(s) \rho_{\mathfrak{A}}(t) 
    - \sigma_{1}^{z}(s) \rho_{\mathfrak{A}}(t) \sigma_{1}^{z}(t) \big) + \Gamma(t,s)^{*} \big( \rho_{\mathfrak{A}}(t) \sigma_{1}^{z}(s) \sigma_{1}^{z}(t) 
    - \sigma_{1}^{z}(t) \rho_{\mathfrak{A}}(t) \sigma_{1}^{z}(s) \big) \Big]
\end{align}
We can then compute the norm of the Liouvillian in the state space as,
\begin{equation}\label{eq:tfim_state}
    \norm{\mathcal{L}_{t}\rho_{\mathfrak{A}}} \leq 4J^{2} \int_{0}^{t} ds\, \| \sigma_{1}^{z}(t) \|_{sp} \| \sigma_{1}^{z}(s) \|_{sp} \Re{\Gamma(t,s)}\,,
\end{equation}
where we have used $\norm{M+N}\leq\norm{M}+\norm{N}$, $\norm{MN}\leq\norm{M}\norm{N}$ and the fact that $\tr{\rho_{\mathfrak{A}}^{2}}\leq1$. This is then used to compute the state bound Eq.~(\ref{eq:qsl-final-b}) in the main text.

In Liouville space, we can write the  Liouvillian operator as,
\begin{equation}
    L_{t} = - J^{2}\int_{0}^{t} ds \Big[ \Gamma(t,s) \big( \sigma_{1}^{z}(t) \sigma_{1}^{z}(s) \otimes I - \sigma_{1}^{z}(s) \otimes \sigma_{1}^{z}(t) \big) +\Gamma(t,s)^{*} \big( I \otimes \sigma_{1}^{z}(t) \sigma_{1}^{z}(s) - \sigma_{1}^{z}(t) \otimes \sigma_{1}^{z}(s) \big) \Big]
\end{equation}
Plugging this expression above into Eq.~(\ref{eq:qsl-final-a}) in the main text, we obtain the Liouville space bound.

In the main text, we evaluated the QSL for this model, where we assumed that the environment $\mathfrak{B}$ consisting of $(N-1)$ spins in a chain of $N$ spins is always in a stationary thermal state. Here, we relax that assumption and compute the reduced state of the environment $\rho_{\mathfrak{B}}(t)$ at all times and compute the environment correlation functions wrt to this non-stationary state. If we relax the assumption of the bath being in stationary state, then the state of the composite system at all times is $\rho(t) = \rho_{\mathfrak{A}}(t)\otimes\rho_{\mathfrak{B}}(t)$. The master equation becomes,
\begin{equation}
    \dot{\rho_{\mathfrak{A}}} = -\lambda^2 \int_0^t ds \,,\sum_{i,j} \Gamma_{i,j}^{'}(t,s)\,
   \Bigl(\mathcal{A}_{i}(t
   )\mathcal{A}_{j}(s)\rho_{\mathfrak{A}} - \mathcal{A}_{j}(s)\rho_{\mathfrak{A}}\mathcal{A}_{i}(t
   )\Bigl) 
    + \text{ h.c}
\end{equation}
where  $\Gamma_{i,j}^{'}= \tr\left\{B_{i}(t) B_{j}(s)\rho_{\mathfrak{B}}(t)\right\}$.
\end{widetext}

\begin{figure}
    \centering
    \includegraphics[scale=0.7]{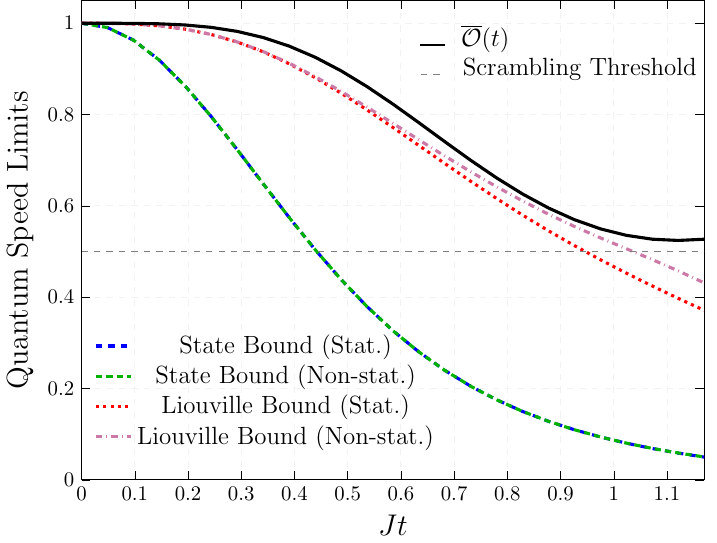}
    \caption{QSL for non-stationary ennvironment for the Ferromagnetic TFIM with $N=10$, $J=0.65$, $h=0.9$, $g=0.4$, $\beta=1.0$}
    \label{fig:full qsl tfim}
\end{figure}

This results in the same form of master equation as Eq.~(\ref{eq:tfim_redfield}) but with a modified environmental correlation function. Using this master equation, we obtain another set of QSL, the results for which are shown in Fig.~\ref{fig:full qsl tfim}. It is observed that for the Liouville space bounds that, indeed the non stationary bath leads us to a tighter speed limit for scrambling but the difference between the stationary thermal and non-stationary environment is pretty insignificant. On the other hand, the state space bounds are loose enough to begin with such that there is no visible difference between the two types of environment when it comes to the QSL.

\begin{figure}
    \centering
    \makebox[0pt][l]{\raisebox{1.5ex}{}}%
    \includegraphics[width=\linewidth]{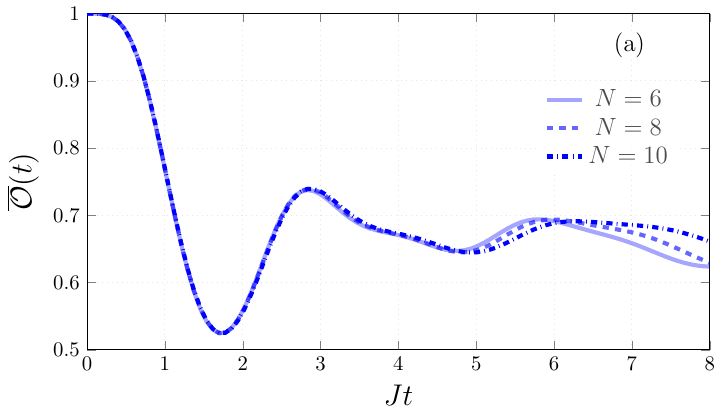}

    \vspace{0.75em}

    \makebox[0pt][l]{\raisebox{1.5ex}{}}%
    \includegraphics[width=\linewidth]{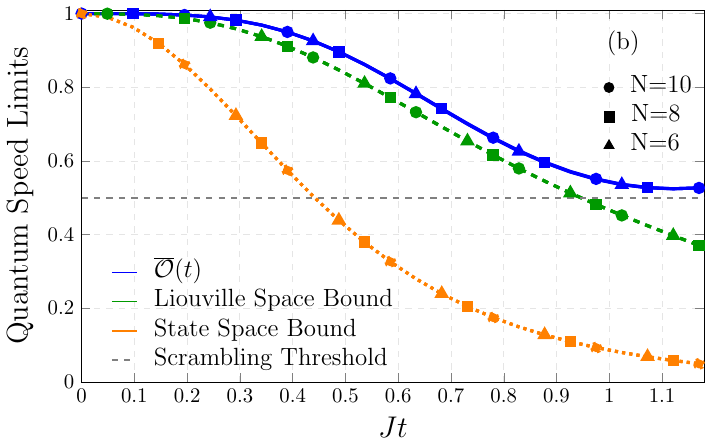}

    \caption{Panel (a): OTOCs for the ferromagnetic non-integrable TFIM with $J=0.65$, $h=0.9$, $g=0.4$, $\beta=1.0$ for different $N$. The OTOCs differ at late times but coincide at early times. Panel (b): QSL for $N=6,8,10$ (same parameters). The QSL bounds and early-time OTOC behavior coincide across $N$.}
    \label{fig:qsl_combined}
\end{figure}

Next, we compute our QSL bounds for different values of $N=6,8,10$. This is shown in Fig.~\ref{fig:qsl_combined}(b). We observe that the $\bar{\mathcal{O}}(t)$ and both the bounds exactly coincide for all $N$s. Figure.~\ref{fig:qsl_combined}(a) shows the long time behavior of $\bar{\mathcal{O}}(t)$ for all $N$s where we find that they differ only in the very late time behavior because of the finite size effects which vary with different sizes of the spin chain.

\begin{figure}
        \centering
        \includegraphics[scale=0.65]{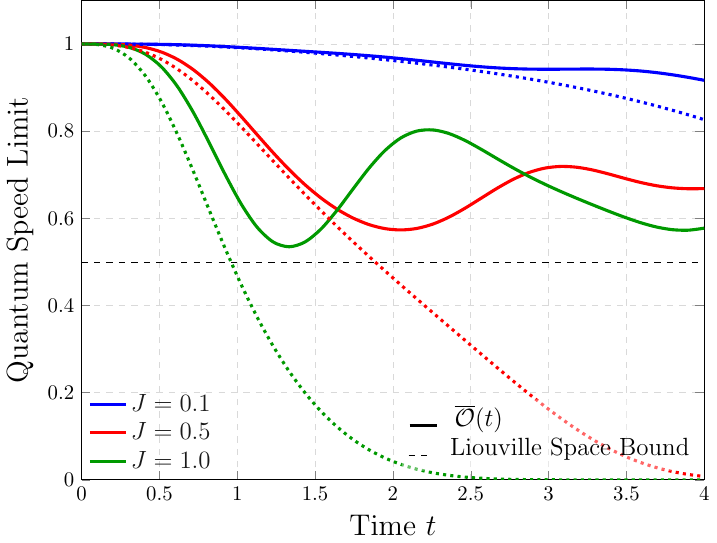}
        \caption{\raggedright 
        Quantum Speed Limit for the ferromagnetic non-integrable TFIM with $N=10$, $h=0.9$, $g=0.4$, $\beta=1.0$ for various $J$.}
        \label{fig:qsl_all_J}
\end{figure}

Figure~\ref{fig:qsl_all_J} illustrates our QSL for different values of coupling $J$, keeping other parameters of the setup constant. Note that the description of the open system as a Redfield master equation is the most appropriate description in the weak coupling limit. This is verified in Fig.~\ref{fig:qsl_all_J}. We find that for small values of $J$, the bound on $\bar{\mathcal{O}}(t)$ is tightest, however for such small values of coupling between the system and environment, the scrambling is slowed down significantly. As we increase $J$, the bound becomes less tight, although it still holds true.


\paragraph{Appendix B: Antiferromagnetic Transverse Field Ising Model}

Finally, we consider the antiferromagnetic non-integrable transverse field Ising model given by the Hamiltonian,
\begin{equation}\label{eq:afm_tfim}
    H = J \sum_{i=1}^{N-1} \sigma_{i}^{z} \sigma_{i+1}^{z} - h \sum_{i=1}^{N} \sigma_{i}^{x} - g \sum_{i=1}^{N} \sigma_{i}^{z}.
\end{equation}
When the longitudinal term proportional to $g$ is a small perturbation to the rest of the Hamiltonian, it is known that under a second order perturbation theory, the total Hamiltonian can be mapped back into an effective integrable antiferromagnetic transverse field Ising model \cite{PhysRevB.68.214406,PhysRevB.92.104306},
\begin{equation}
    H_{\text{eff}}\approx(1-g^{2})\sum_{i}\sigma_{i}^{z}\sigma_{i+1}^{z}-\sum_{i}\sigma_{i}^{x}
\end{equation}
where we have set $J=1.0$ and $h=0.5$.

Since this model is analytically solvable, we can compute the $ZZ$ two point correlation function at the edge site in the ground state of the Hamiltonian which is,
\begin{equation}\label{eq:analytic_two_pt}
    \langle\sigma^{z}(t)\sigma^{z}(0)\rangle = 4\sum_k\frac{\psi^2_k}{\Gamma^2_k}\, \exp{-i\Gamma_k t}
\end{equation}
where $ \psi_k=\sqrt{2/N}\sin k $
\begin{equation}
    \Gamma_k=2\sqrt{(1-g)^{2}+1+2(1-g)\cos k}
\end{equation}
The allowed wave-vectors in the summation are: $k=n\pi/N$, with $n=1,2,3,\cdots, N$. In the large $N$ limit, we can convert the summation into an integral over the continuum modes $k$
\begin{eqnarray}
    \sum_kG_k&\to&\frac{N}{\pi}\int_0^\pi dk\: G(k).
\end{eqnarray}
One can then plug Eq.~(\ref{eq:analytic_two_pt}) into the environmental two point correlation function in Eq.~(\ref{eq:tfim_redfield}) of the main text and compute the corresponding QSLs analytically.

\end{document}